\documentstyle[twoside,fleqn,espcrc2,epsf]{article}

\newcommand{\beq}{\begin{equation}}

\newcommand{\eeq}{\end{equation}}
\newcommand{\bea}{\begin{eqnarray}}
\newcommand{\eea}{\end{eqnarray}}
\newcommand{\F}{\Phi}
\newcommand{\f}{\phi}
\newcommand{\vf}{\varphi}
\newcommand{\Q}{\tilde{Q}_{_L}}
\newcommand{\q}{\tilde{q}_{_R}}
\newcommand{\Lp}{\tilde{L}_{_L}}
\newcommand{\lp}{\tilde{l}_{_R}}
\newcommand{\e}{{\cal E}_\omega}

\newcommand{\AmS}{{\protect\the\textfont2
  A\kern-.1667em\lower.5ex\hbox{M}\kern-.125emS}}

\title{ Q-balls in the MSSM
}

\author{Alexander Kusenko\address{Theory Division, CERN,
1211 Geneva, Switzerland}%
        \thanks{email address: Alexander.Kusenko@cern.ch}
}
       
\begin{document}

\begin{abstract}
Q-balls are generically present in models with softly broken low-energy
supersymmetry.  We discuss the properties of these 
non-topological solitons, which can precipitate a new kind of first-order
phase transition in the early Universe and have other important 
implications for cosmology and experiment. 
\end{abstract}

\maketitle

Non-topological solitons have been studied in a variety of models 
\cite{tdlee}.  One particular type of such solitons, Q-balls
\cite{tdlee,coleman1}, are present in the spectrum of a theory, 
whenever it satisfies the conditions outlined below.  This class of theories
includes the MSSM and other supersymmetric generalizations of the Standard
Model.   Surprisingly, the rich phenomenology associated with the solitonic
sector of the MSSM has not been explored until very 
recently~\cite{ak_mssm,ak_pt}. 

As a toy model, let us consider a field theory with a scalar potential
$U(\vf) $  that has a global minimum $U(0)=0$ at $\vf=0$.
Let $U(\vf)$ have an unbroken global\footnote{
Q-balls associated with a local symmetry have been constructed 
\cite{l}.  An important qualitative difference is that, in the case of a
local symmetry, there is an upper limit on the charge of a stable Q-ball.} 
U(1) symmetry at the origin, $\vf=0$.  And let the scalar field $\vf$ have
a unit charge with respect to this U(1).

The charge of some field configuration $\vf(x,t)$ is  
 
\beq
Q= \frac{1}{2i} \int \vf^* \stackrel{\leftrightarrow}{\partial}_t  
\vf \, d^3x . 
\label{Qt}
\eeq
Since a  trivial configuration $\vf(x)\equiv 0$ has zero charge, the
solution that minimizes the energy, 
 
\beq
E=\int d^3x \ \left [ \frac{1}{2} |\dot{\vf}|^2+
\frac{1}{2} |\nabla \vf|^2 
+U(\vf) \right], 
\label{e}
\eeq
and has a given charge $Q>0$, must differ from zero in some (finite)
domain.  This is a Q-ball.   It is a time-dependent solution, more
precisely it has a time-dependent phase. However, all physical quantities
are time-independent.  Of course, we have not proven that such a 
``lump'' is finite, or that is has a lesser energy than the collection of
free particles with the same charge; neither is true for a general
potential.  A {\it finite-size } Q-ball is a minimum of energy and is
stable with respect to decay into free $\vf$-particles if 

\beq
U(\vf) \left/ \vf^2 \right. = {\rm min},
\ \ {\rm for} \ 
\vf=\vf_0>0 
\label{condmin}
\eeq

One can show that the equations of motion for a Q-ball in 3+1 dimensions
are equivalent to those for the bounce associated with tunneling 
in 3 Euclidean dimensions
in an effective potential $\hat{U}_\omega
(\vf)= U(\vf) - (1/2) \omega^2 \vf^2$, where $\omega$ is such that it
extremizes~\cite{ak_qb}

\beq
\e = S_3(\omega) +\omega Q. 
\label{Ew}
\eeq
Here $S_3(\omega)$ is the three-dimensional Euclidean action of the bounce
in the potential $\hat{U}_\omega (\vf)$ shown in Figure~1.  The Q-ball
solution has the form 

\beq
\vf(x,t) = e^{i\omega t} \bar{\vf}(x),
\eeq
where $\bar{\vf}(x)$ is the bounce. 

The analogy with tunneling clarifies the meaning of the condition 
(\ref{condmin}), which simply requires that there exist a value of 
$\omega$, for which $\hat{U}_\omega (\vf)$ is negative for some value of 
$\vf=\vf_0 \neq 0$ separated from the false vacuum by a barrier. 
This condition ensures the  existence of a bounce.  (Clearly, the
bounce does not exist if $\hat{U}_\omega (\vf) \ge 0$ for all $\vf$ because
there is nowhere to tunnel.)  

\begin{figure}
\setlength{\epsfxsize}{3in}
\setlength{\epsfysize}{2.4in}
\centerline{\epsfbox{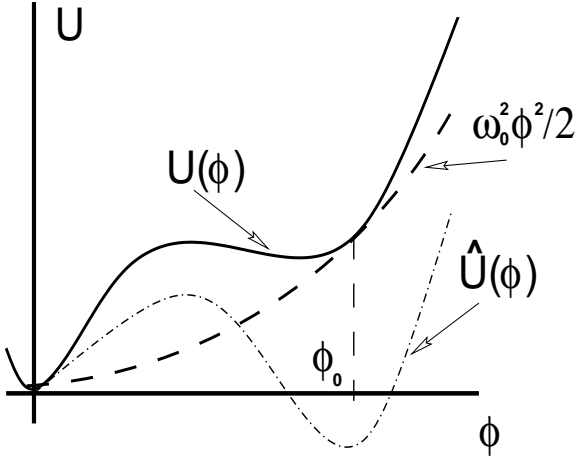}}
\caption{
The scalar potential {$U(\vf)$} (solid line) and the effective potential 
{$\hat{U}_\omega (\vf)$} (dash-dotted line) for some value of {$\omega$}. 
As charge increases, {$\omega$} decreases approaching {$\omega_0$}, the
coefficient of a parabola tangential to  {$U(\vf)$} (dashed line).    
} 
\label{fig1}
\end{figure}

In the true vacuum, there is a minimal value $\omega_0$, so that only for  
$\omega>\omega_0$, $\hat{U}_\omega (\vf)$ is somewhere negative (see
Figure~1).  If one considers a Q-ball in a metastable false vacuum, then  
$\omega_0=0$.  The mass of the $\vf$ particle is the upper bound on
$\omega$ in either case. Large values of $\omega$ correspond to small
charges~\cite{ak_qb}.  As $Q \rightarrow \infty$, $\omega \rightarrow 
\omega_0$.  In this case, the effective potential $\hat{U}_\omega (\vf)$
has two nearly-degenerate minima; and one can apply the thin-wall
approximation to calculate the Q-ball energy~\cite{coleman1}.  For smaller 
charges, the thin-wall approximation breaks down, and one has to resort to
other methods~\cite{ak_qb}. 

The above discussion can be generalized to the case of several fields, 
$\vf_k$, with different charges, $q_k$~\cite{ak_mssm}.  Then the Q-ball is
a solution of the form 

\beq
\vf_k(x,t) = e^{iq_k \omega t} \vf_k(x),
\label{tsol}
\eeq
where $\vf(x)$ is again a three-dimensional bounce associated with
tunneling in the potential 

\beq
\hat{U}_\omega (\vf) = U(\vf)\ - \ \frac{1}{2} \omega^2 \, 
\sum_k q_k^2 \, |\vf_k|^2. 
\label{Uhat}
\eeq
As before, the value of $\omega$ is found by minimizing $\e$ in equation 
(\ref{Ew}).  The bounce, and, therefore, the Q-ball, exists if 

\begin{eqnarray}
\mu^2 & = & 
2 U(\vf) \left/ \left (\sum_k q_k \vf_{k,0}^2 \right ) \right. = {\rm min},
\ \nonumber \\ & & {\rm for} \ |\vec{\vf}_0|^2 > 0.
\label{condmin1}
\end{eqnarray}
Of course, evaluating the soliton mass is as difficult in this case as is
calculating the bounce in a potential that depends on several fields.  
However, one can use, {\it e.\,g}., the method of Ref.~\cite{ak_tunn} to
compute $S_3(\omega)$ numerically.

In the MSSM, the superpartners of quarks and leptons carry a non-zero
baryon and lepton number.  In addition, the structure of the scalar
potential is such that the condition (\ref{condmin1}) is automatically
satisfied unless some Yukawa couplings and some soft supersymmetry breaking
terms are zero~\cite{ak_mssm}.  Therefore, Q-balls associated with baryon 
(B) and lepton (L) number conservation are generically present in the
MSSM. 

Every supersymmetric generalization of the Standard Model must have Yukawa
couplings  of the Higgs fields $H_1 $ and $H_2$ to quarks and leptons which
arise from the superpotential of the form 

\beq
W=y H_2 \F \f + \tilde{\mu} H_1 H_2 +...
\label{sptn}
\eeq
Here $\F$ stands for either a left-handed quark ($\Q$), or a 
lepton ($\Lp$) superfield, and $\f$ denotes $\q$ or $\lp$, respectively.  
The corresponding scalar potential must, therefore, have cubic terms of the
form  $y \tilde{\mu} H_2 \F \f$.  In addition, there are soft supersymmetry
breaking terms of the form $y A H_1 \F \f$.  This is a generic feature of
all SSM.  

For squarks and sleptons, there are several abelian symmetries that are
suitable for building Q-balls.  These are $U(1)_{B}$, $U(1)_{{L_i}}$ and
$U(1)_{E}$,  associated with the conservation of baryon number, three types
of lepton numbers, and the electric charge.  

In the MSSM, Q-balls are allowed, therefore, to have a baryon number, a 
lepton number, and an electric charge.  As a toy model, one can consider a
potential for the Higgs field, $H$, and a pair of sleptons, $\Lp$ and
$\lp$, with a scalar potential  

\begin{eqnarray}
U & = & m_{_H}^2 |H|^2+m_{_L}^2 |\Lp|^2+ m_{l}^2|\lp|^2  \nonumber \\ 
& & - y A (H \Lp^* \lp +c.c.) \nonumber \\ 
& & + y^2 (|H^2 \Lp^2| + |H^2 \lp^2| + |\Lp^2 \lp^2|) \nonumber \\ 
& & + V_{_D}, 
\label{toy}
\end{eqnarray}
where $V_{_D}= (g_1^2/8) [|H|^2- |\Lp|^2]^2 +(g_2^2/8) [|H|^2+|\Lp|^2-2
|\lp|^2]^2$ 
is the contribution of the gauge the $D$-terms.   For simplicity, we 
neglected the Higgs VEV.   Nevertheless, this
toy model is instructive because it allows for some non-topological
solitons with the same quantum numbers as those in the MSSM.  The potential
is invariant under the global $U(1)_{_L}$ symmetry 
($\Lp \rightarrow \exp \{i \theta \}\Lp$ and $\lp \rightarrow \exp \{i
\theta \} \lp$) associated with the lepton 
number conservation.  Both $\Lp$ and $\lp$ have a unit charge with respect
to this $U(1)$, while the Higgs field is $U(1)_{_L}$ invariant. 

It is convenient to write

\beq
\left \{ \begin{array}{lll}
H &=& F \ sin\xi \\
\Lp &=& F \ \cos\xi \ \sin \theta \\
\lp &=& F \ \cos\xi \ \cos \theta 
\end{array} 
\right.
\label{fields}
\eeq 
The condition (\ref{condmin1}) is satisfied, and a Q-ball with mass 
$M_{_Q}=\mu Q$ exists,  if $\mu^2$ is minimized at some value of $F\neq 0$.

\begin{eqnarray}
\mu^2 & = &  \frac{2U}{|\Lp|^2 + |\lp|^2} = \nonumber \\
& & \frac{1}{\cos^2 \xi} [\gamma_2(m_i^2,\xi.\theta) - \nonumber \\
& & y A \gamma_3(\xi,\theta) \, F \ + \  \gamma_4 (\xi,\theta) \, F^2], 
\label{cond_gammas} 
\end{eqnarray}
where $\gamma_2$ and $\gamma_4$  are non-negative functions of masses and
mixing angles, $\gamma_3= \cos^2\xi \,\sin \xi \, \sin(2\theta) $.  
The minimum of $\mu^2$ in (\ref{cond_gammas}) is achieved 
at $F \neq 0 \ $ if $\ y A \neq 0$.  The origin $F=0$ is not a local
minimum.   Therefore, in our toy model, $L$ balls exist no matter how small
the  tri-linear couplings might be, as long as they are non-zero.  The same
is true of the baryonic balls built of squarks.  

Of course, in the full MSSM there can be other fields that carry the same
charge.  Therefore, the local minimum of energy corresponding 
to a particular set of fields may not be the global minimum.  For example, 
an electrically  neutral selectron  $L$ ball,
$\{H,\tilde{e}_{_L},\tilde{e}_{_R} \}$, will be in competition with a 
sneutrino ball, $\{H,\tilde{\nu}_{_L},\tilde{\nu}_{_R} \}$.  However, since
the origin is not a local minimum of (\ref{cond_gammas}) for $y A \neq 0$, 
there is always a stable Q-ball with a given lepton (baryon)
number.

Supersymmetric B-balls and L-balls can form in the early Universe in
several ways.  First, they can be produced via Kibble mechanism in the
course of a phase transition.  Second, small-charge Q-balls (``Q-beads'')
can be pair-produced at finite temperature.  Third, thermal fluctuations of
a baryonic and leptonic charge can, under some conditions, form a Q-ball. 
Finally, a process of a gradual charge accretion, similar to
nucleosynthesis, can take place.  If the latter occurs in the false vacuum,
critical Q-balls produced via solitosynthesis can destabilize the
metastable vacuum even if the tunneling probability is negligibly small.  

\begin{figure}
\setlength{\epsfxsize}{3in}
\setlength{\epsfysize}{2.4in}
\centerline{\epsfbox{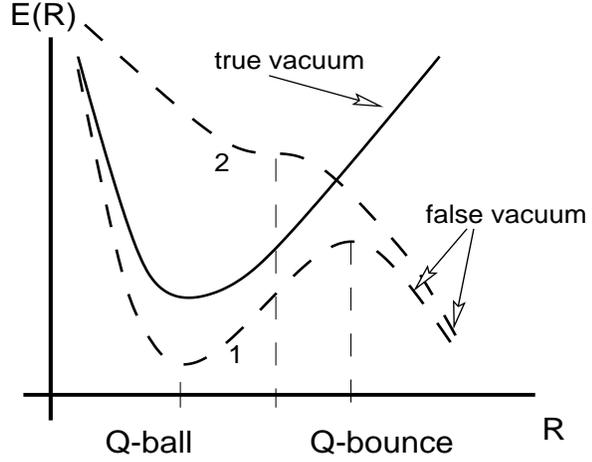}}
\caption{
Energy (mass) of a soliton as a function of its size.  In the true vacuum, 
Q-ball is the global minimum of energy (solid curve).  In the false vacuum,
if the charge is less than some critical value, there are two solutions: a
``stable'' Q-ball, and an unstable ``Q-bounce'' (dashed curve 1) .  In the
case of a critical charge (curve 2), there is only one solution, which is
unstable. 
} 
\label{fig2}
\end{figure}

Indeed, the potential energy in the Q-ball interior is positive in the case
of a true vacuum, but negative if the system is in the metastable false
vacuum. In either case, it grows as the third power of the Q-ball radius
$R$ (assuming $\vf_0=const$).   
The positive contribution of the time derivative to the soliton mass can be
written as  $Q^2/\int \vf^2(x)d^3x \propto R^{-3}$, and the gradient
surface energy scales as $R^2$.  In the true vacuum, all three
contributions are positive and the Q-ball is the absolute minimum of energy
(Figure~2).  However, in the false vacuum, the potential energy inside the
Q-ball is negative and goes as $\propto -R^3$.  As shown in Figure~2, 
for small charge $Q$, there are two stationary points, the minimum and the 
maximum.  The former corresponds to a Q-ball (which is, roughly, as stable
as the false vacuum is), while the latter is a critical bubble of the true
vacuum with a non-zero charge.  

There is a critical value of charge $Q=Q_c$, for which the only stationary
point is unstable.  If formed, such an unstable solution will expand
filling the space with the true vacuum phase.  

This allows for a new type of a first-order phase transition at finite
temperature.  Q-ball is the minimum of energy $E$ but is also a state of
small entropy $S$  (because it is an extended object).  For high
temperature it is not a minimum of free energy $F=E-TS$, because of the  
$TS$ term.  
At some small enough temperature, however, the second term in $F$ is no
longer important, and the gain from minimizing $E$ overwhelms the loss from
lowering the entropy.  At that point, Q-ball is the minimum of free
energy.  Under some conditions, a copious production of Q-balls can take
place~\cite{ak_pt}. 

As the charge of a growing Q-ball reaches the critical value $Q_c$, the phase
transition takes place.  It was shown~\cite{ak_pt} that a critical Q-ball with
charge as small as $10^2$ -- $10^3$ can destabilize a false vacuum, which
would otherwise be stable on the time scales of order the present age of the
Universe.  

Tunneling is a very improbable event because one has to wait for many
random quanta to fall into the right places and form an extended object, 
a critical bubble.  One could use an analogy with building a house by
throwing bricks from a helicopter: it is highly unlikely that a random 
pile produced in such experiment will coincide with the designer's
blueprint.  There is, however, a better way to construct a house, that is 
building it brick by brick.  This is precisely the option the Q-balls allow
with respect to building a critical bubble.  The role of charge
conservation is to keep the small bubbles from collapsing while the charge 
accretion takes place.   Once a critical size is reached, it becomes
energetically preferable for the bubble interior to expand (see Figure~2). 

An example of such phase transition in the MSSM, from a charge-breaking
minimum to the standard vacuum, was discussed in Ref.~\cite{ak_pt}.  
Color and charge breaking (CCB) minima have been a subject of considerable
attention because of their implications for the allowed regions of the MSSM
parameters (see, {\it e.\,g.}, Ref.~\cite{kls} and references therein). 
The CCB minima with negative energy are allowed if ($i$) the evolution of the
early Universe leads to the ``right'' vacuum, which ($ii$) can survive for 
some 10 billion years or more before decaying through tunneling. 
The possibility of a phase transition precipitated by
solitosynthesis~\cite{ak_pt} can alter some of the corresponding bounds.  

Another interesting question is the effect of B- and L-balls on
nucleosynthesis.  Suppose such solitons are formed at some high
temperature $\sim 100$ GeV.  Then they will evaporate by emitting the
baryonic (leptonic) charge in terms of quarks and leptons.  
In general, the  evaporation of a large Q-ball into fermions is a slow
process because its rate is proportional to the surface area, while the 
total charge is proportional to the soliton volume~\cite{ccgm}. 
Those $B$ and $L$ solitons that decay at temperatures $T$ above 1~GeV, 
probably, have no observable consequences.  However, a remarkable 
transformation can take place for a Q-ball that survived to a temperature of
order $\Lambda_{QCD}$.  We recall that the interior of a large evaporating
Q-ball is populated with a high density of quarks that fill up 
the Dirac sea~\cite{ccgm}.  If the Q-ball survives to temperatures
below $\Lambda_{QCD}$, then the population of quarks fostered inside the 
sparticle ball can remain bound, now by the QCD forces, even after the 
sparticle structure, which kept them together originally, disappears.  
At $T \gg \Lambda_{QCD}$, such a conglomerate of nuclear matter would
thermalize without a trace.  However, at lower temperatures, heavy nuclei
can form as vestiges of sparticle Q-balls.

In principle, this allows for a highly non-standard synthesis of heavy
nuclei in the early Universe, such that they are already present at the
time $t\sim 1$~s, when the standard nucleosynthesis is supposed to
commence.  Fission of heavy nuclei can also be the source of additional
lighter elements, in particular, $^4He$, which are copiously produced in
nuclear decays.  Of course, it may be difficult to reproduce 
the observed element abundances in such scenario.  A more conservative
approach can yield the {\it status quo} constraints on the MSSM
parameters from the non-interference of Q-balls with the nucleosynthesis. 

Q-balls with lifetime longer than 1~second are probably disallowed, at
least if they can be produced in substantial quantities. Their decay
products can cause an unacceptable increase in entropy, or disturb
the spectrum of the microwave background radiation. 

Although there is no classical lower limit on the charge of a stable Q-ball,
it is unclear whether small non-topological solitons can be produced in a
collider experiment because they are extended objects, whose size is large
in comparison to their Compton wavelength.  A vaguely related question of
whether a bounce can be created in a collider experiment 
has been the subject of much study and debate~\cite{mattis}. 

To summarize, Q-balls are generically present in all phenomenologically 
acceptable supersymmetric extensions of the Standard Model.  Their
production in the early Universe can cause a number of interesting
phenomena.  In particular, the synthesis of Q-balls in the false vacuum,
which proceeds via charge accretion, can precipitate an otherwise
impossible first-order phase transition.

\end{document}